\documentstyle[eqsecnum,aps,preprint,prl]{revtex}

\title{Possible Spin Polarization in
a One-Dimensional Electron Gas}
\author{K.~J. Thomas, J.~T. Nicholls, M.~Y. Simmons,
M. Pepper, D.~R. Mace, and D.~A. Ritchie} 
\address{Cavendish Laboratory, Madingley Road, 
Cambridge CB3 0HE, United Kingdom}

\date{\today}
\begin{document}
\maketitle

\begin{abstract}
In zero magnetic field, conductance measurements of clean 
one-dimensional (1D) constrictions defined
in GaAs/AlGaAs heterostructures show 
twenty-six quantized ballistic plateaux,
as well as a structure close to $0.7(2e^2/h)$.
In an in-plane magnetic field all the 1D subbands show Zeeman splitting
and in the wide channel limit the $g$-factor is $\mid g \mid = 0.4$, 
close to that of bulk GaAs.
For the last subband spin-splitting originates 
from the structure at $0.7(2e^2/h)$,
indicating spin polarization at $B=0$.
The measured enhancement of the $g$-factor as the subbands are depopulated
suggests that the ``0.7 structure'' 
is induced by electron-electron interactions.

\end{abstract}
\pacs{71.70.Ej, 71.25.Jd, 73.40-c, 73.20Dx} 

The application of a negative 
voltage to lithographically defined gates 
over a GaAs-AlGaAs heterostructure
allows the underlying two-dimensional 
electron gas (2DEG) to 
be electrostatically squeezed into 
a particular shape.\cite{thornton86} 
This has allowed the study of one-dimensional 
transport phenomena,\cite{BeenRev} 
where if the mean free path is larger than the 
effective channel length, it is possible to observe
ballistic one-dimensional (1D) conductance plateaux.\cite{Wharam88,VanWees88}

Interaction effects may be significant 
in a clean one-dimensional single-mode electron gas,
giving rise, for example,
to new crystal\cite{glazman92,schulz93}
and liquid states.\cite{haldane81}
For long-range $1/r$ interactions 
between electrons a 1D Wigner crystal 
is expected\cite{glazman92} to occur when the 1D carrier density 
is much less than the $(Bohr\ radius)^{-1}$. 
For short-range interactions 
a clean 1D system may be modeled as
a Tomonaga-Luttinger (TL) liquid
where it is predicted\cite{kane92c} that the conductance 
will be renormalized to $G=K(2e^2/h)$,
with a parameter $K > 1$ for attractive interactions, 
$K < 1$ for repulsive interactions, 
and $K=1$ for a non-interacting electron gas.
Other theories\cite{maslov95,safi95,pmarenko95} 
suggest that conductance renormalization will not occur
because the measured contact resistance
is determined by the injection of
non-interacting electrons into the wire.

With the exception of recent results,\cite{tarucha95}
phenomena observed at zero magnetic field 
in clean 1D GaAs wires have been 
interpreted within a single-electron picture.
Using a modified TL liquid theory\cite{ogata94} 
that accounts for disorder scattering, 
Tarucha {\it et al.}\cite{tarucha95}
measured an interaction parameter $K \approx 0.7$ 
from temperature studies of wires longer than 2~$\mu$m.
However this is not supported by the presence of
renormalized conductance quantization.
Experimentally, residual impurity 
scattering and weak resonance effects
make it difficult to interpret 
small changes in $G$ as interaction effects.

In this paper we present two pieces of experimental 
evidence that suggest that interaction 
effects are important in clean split-gate devices.
First, as the number of 1D subbands decreases,  
we have measured an enhancement of the in-plane
electron $g$-factor over its bulk GaAs value.
Second, in zero magnetic field we have observed 
reproducible structure at approximately $0.7(2e^2/h)$.
We show that this {\em 0.7 structure} is an intrinsic property of a 
1D channel at low densities,
and that its origin could be related to spin.

Previously from measurements\cite{patel91a} of 
the transconductance $dG/dV_g$ 
we have determined $g_{\parallel}$, 
the in-plane $g$-factor when the magnetic field 
$\vec{B}$ is applied parallel to the current 
$\vec{j}$ through a ballistic 1D constriction.
A zero in $dG/dV_g$ corresponds to a conductance plateau,
whereas a peak corresponds to the step region between plateaux.
The magnitude (but not the sign) 
of $g_{\parallel}$ was determined 
from a comparison of the splitting of a transconductance peak
in an in-plane magnetic field with that 
induced by an applied source-drain voltage $V_{sd}$.
When the gate voltage separation 
of given peaks in $dG/dV_{g}$ 
of the two measurements are the same,
the $g$-factor is determined by 
equating the two energy scales,\cite{patel91a}
\begin{equation}
eV_{sd} = 2 g \mu_B B S,
\label{energy}
\end{equation}
where $\mu_B$ is the Bohr magneton and $S=1/2$.
When two and three 1D subbands are occupied, 
the measured\cite{patel91a} $g_{\parallel}$-factors
were 1.08 and 1.04, respectively. 
In this earlier experiment it was 
not possible to measure $g_{\perp}$
(in-plane field $\vec{B}$ applied perpendicular to $\vec{j}$).

\begin{figure}
\caption{Temperature dependence of the quantized
plateau at $2e^2/h$ and the structure at $0.7(2e^2/h)$.}
\label{f:4}
\end{figure}
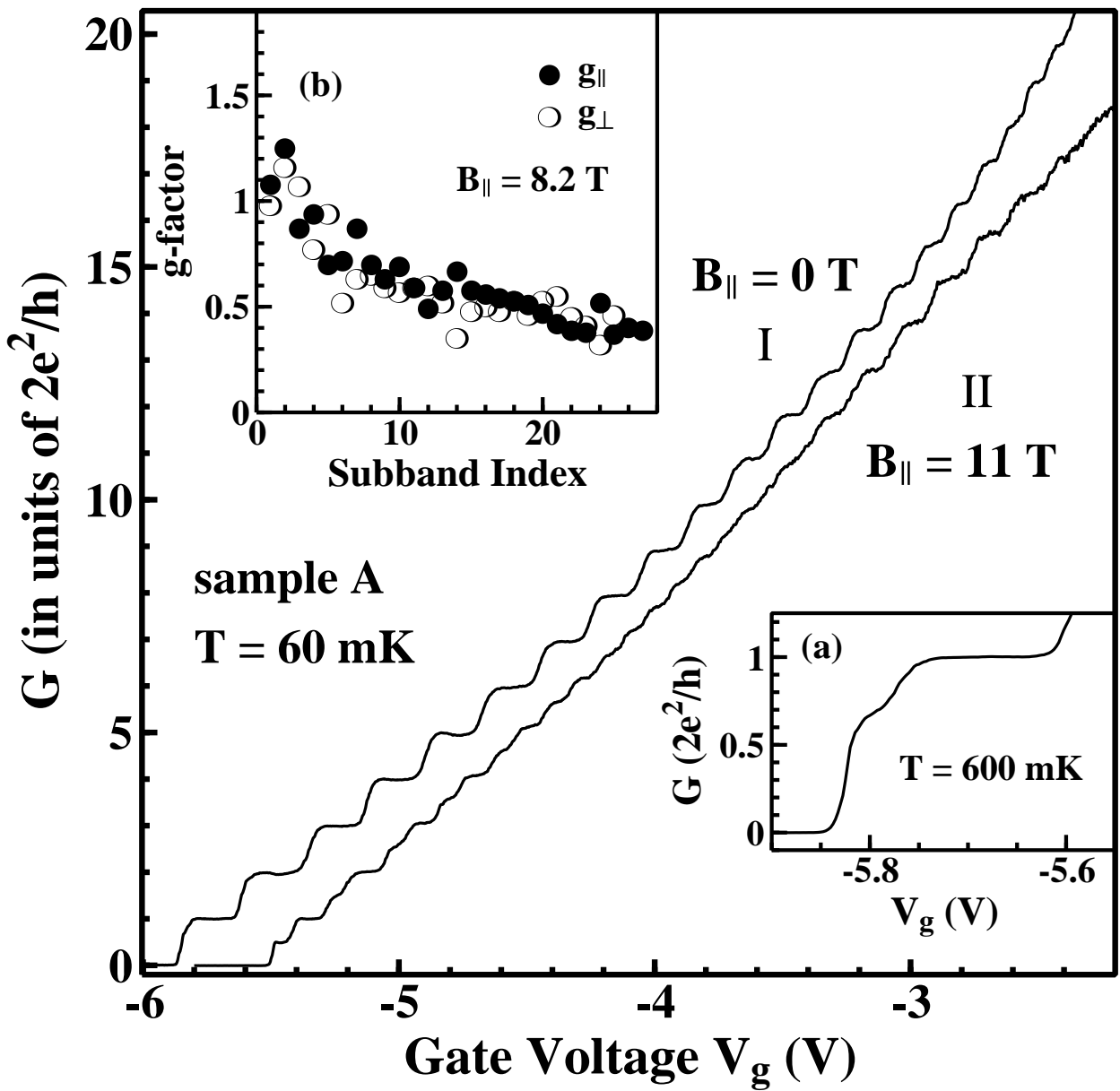
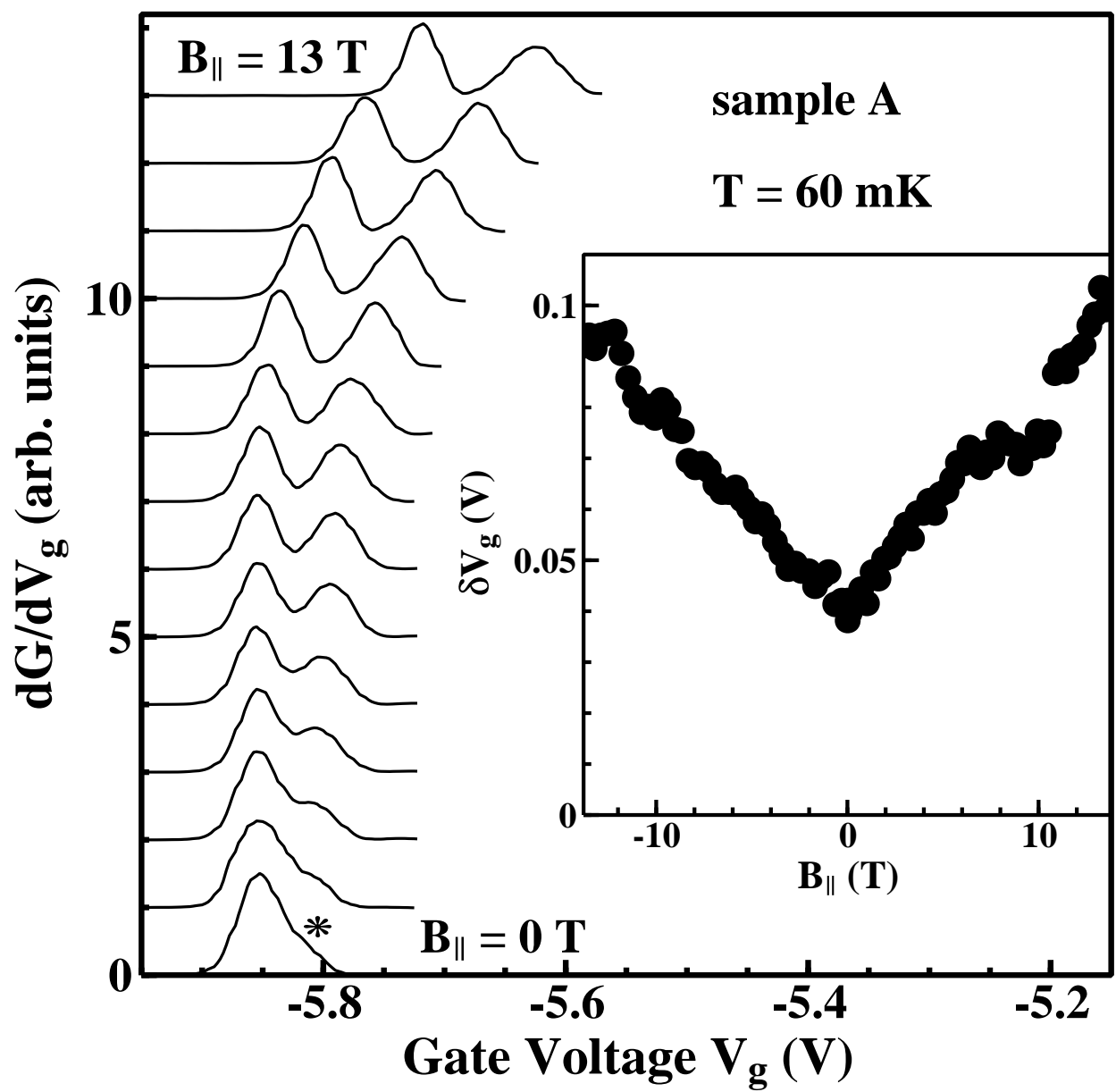
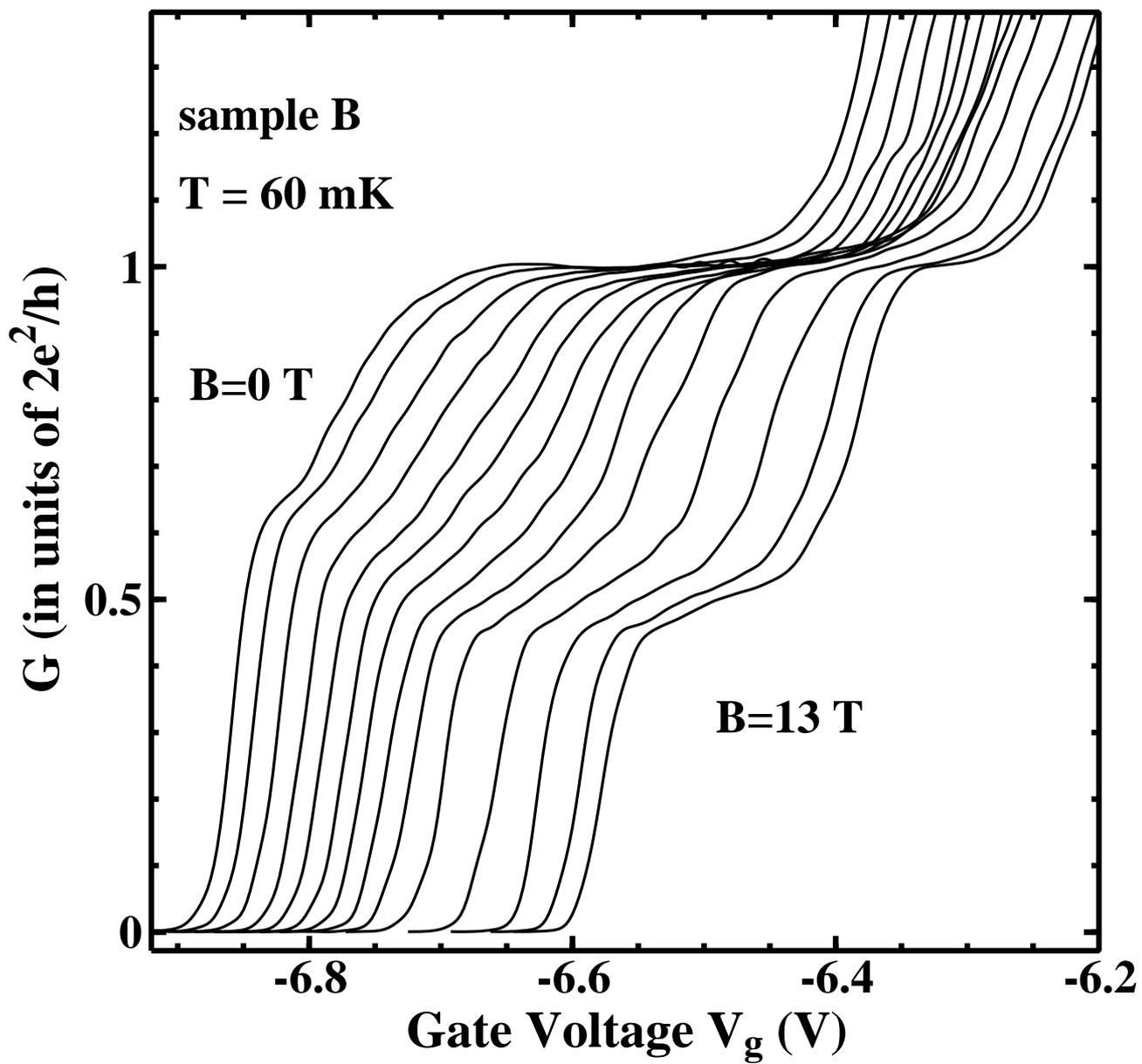
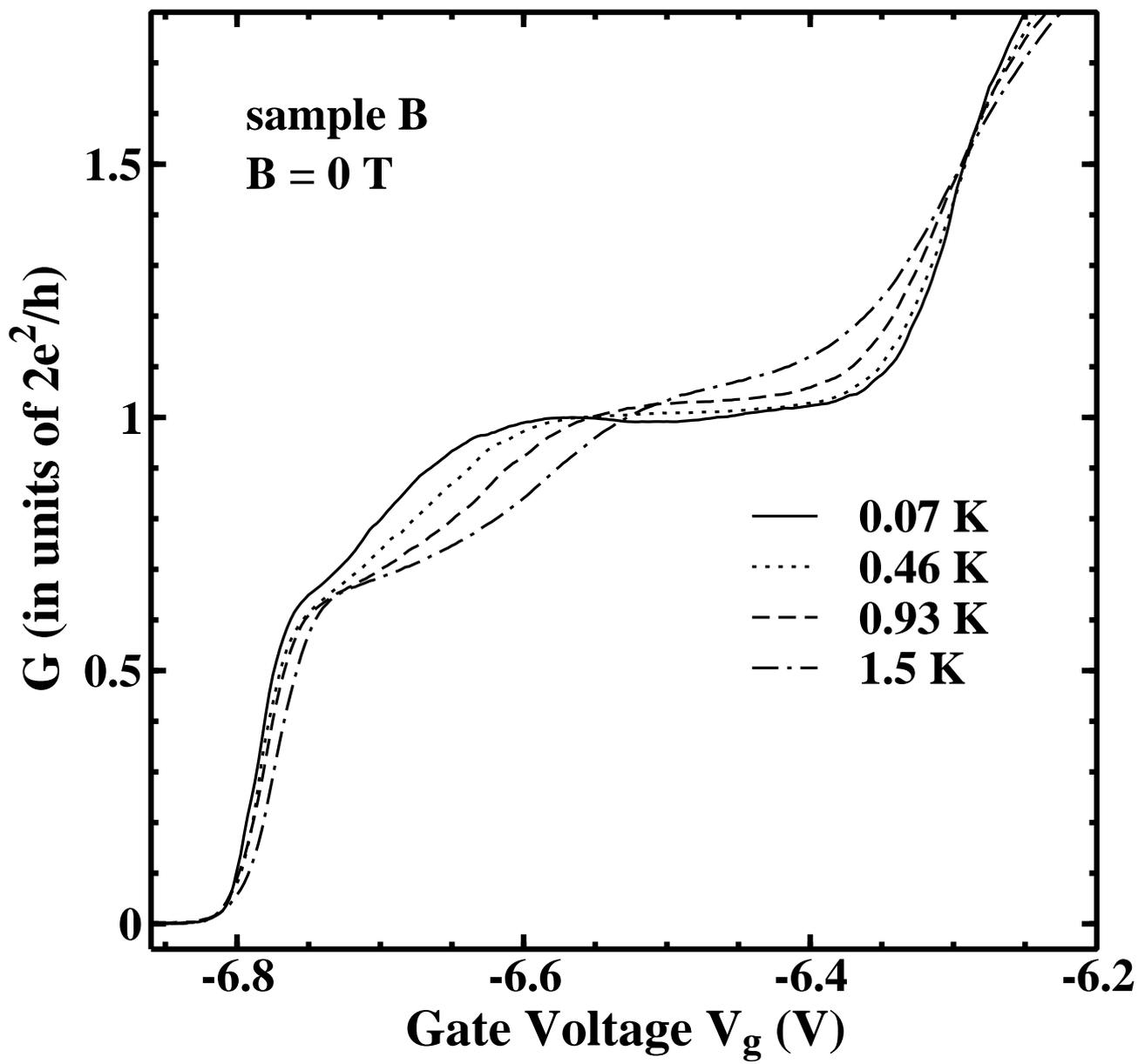

We have investigated ballistic 1D 
constrictions defined in high mobility 2DEGs, 
formed at a modulation-doped GaAs/Al$_{0.33}$Ga$_{0.67}$As heterostructure.
Sample A (Figs.~\ref{f:1} and~\ref{f:2}) is a split-gate device 
of lithographic width $W=0.75~\mu$m and length $L=0.4~\mu$m,
defined above a 2DEG of depth $2770$~\AA,
which after illumination with a 
red light-emitting diode has a carrier density of 
$n_s=1.8 \times 10^{11}$~cm$^{-2}$ 
and a low-temperature mobility
of $\mu=4.5 \times 10^6$ cm$^2$/Vs.
Sample B (Figs.~\ref{f:3} and~\ref{f:4}) is a device
with $W=0.95~\mu$m and $L=0.4~\mu$m,
defined above a 2DEG of depth $3100$~\AA,
which has $n_s = 1.4 \times 10^{11}$~cm$^{-2}$ 
and $\mu = 3.5 \times 10^6$ cm$^2$/Vs after illumination.
Similar 1D constrictions have shown\cite{thomas95} an
absence of resonant structures on the quantized conductance plateaux,
demonstrating the lack of potential 
fluctuations within the 1D constriction.
The 1D subband structure was also probed\cite{thomas95}
using a source-drain voltage of up to $V_{sd}=4$~mV.
For all subbands the splitting of the transconductance 
peaks was linear in $V_{sd}$,
indicating that $V_{sd}$ does not perturb
the electrostatic confinement potential within the constriction.
We shall rely on this result when we use Eq.~\ref{energy} to 
measure both $g_{\parallel}$ and $g_{\perp}$ for all twenty-six 1D subbands.

Low temperature measurements of the two-terminal conductance, 
$G(V_g)=dI/dV$, were performed using an 
excitation voltage of 10~$\mu$V at a frequency of 71~Hz.
Measurements in an in-plane magnetic field were carried
out with the field applied either parallel ($B_{\parallel}$)
or perpendicular ($B_{\perp}$) to the current $j$ 
through the constriction.
The results presented here are qualitatively the same 
for both field orientations.
To check for an out-of-plane magnetic 
field component due to misalignment, 
we monitored the Hall voltage; 
from such measurements we were able to align 
the samples to better than $1^{\circ}$. 
All the results presented in this paper were 
reproducible on different sample cooldowns,
and have been observed in a variety of devices 
fabricated on different wafers.
The bulk 2DEG resistance changes with $B$,
and so conductance sweeps have been corrected by choosing a series
resistance that will match the 
last spin-degenerate plateau to 
the quantized value of $2e^2/h$.

Trace I in Fig.~\ref{f:1} shows the gate characteristics 
$G(V_{g})$ of sample A in zero magnetic field.
With decreasing gate voltage $V_g$ the constriction is
narrowed and a conductance step of $2e^2/h$
is observed each time a spin-degenerate 
1D subband is depopulated.\cite{Wharam88,VanWees88}
In addition to the last plateau at $2e^2/h$,
trace I shows weak structure close to $0.7(2e^2/h)$;
inset (a) in Fig.~\ref{f:1} shows detail of this 
{\em 0.7 structure} at 600~mK.
We have observed this feature in many 1D constrictions
and previously we have commented\cite{patel91b} on 
this reproducible structure,
both in our devices and in others.

We shall first present the magnetic field 
properties of the higher conductance plateaux,
where the application of a magnetic field in the plane of the 2DEG
lifts the spin degeneracy of the 1D subbands.\cite{Wharam88}
Trace II in Fig.~\ref{f:1} shows the gate characteristics 
obtained in a parallel field of $B_{\parallel}=11$~T.
For $V_g < -4$~V additional spin-split plateaux
are interleaved between those observed at $B_{\parallel}=0$. 
For $V_g > -4$~V the Zeeman energy is 
comparable to the subband spacings and 
both sets of spin-split plateaux cannot be easily resolved.
Inset (b) of Fig.~\ref{f:1} shows $g_{\parallel}$ 
and $g_{\perp}$ for all twenty-six subbands,
measured using Eq.~\ref{energy} at 8.2~T;
similar results were obtained at 12~T. 
Three features are clear: 
First, there is little in-plane anisotropy of the $g$-factor.
Second, when the constriction is very wide  
$g_{\parallel} \approx g_{\perp} \approx 0.4$,
close to the value $\mid g \mid = 0.44$ of bulk GaAs.\cite{white72}
Third, as the subband index decreases, 
$g_{\parallel}$ and $g_{\perp}$ increase.

When the channel is just defined,
it is wide with a carrier density equal to that of the bulk 2DEG and 
confinement which can be described by a square-well potential.
In this limit the channel can be considered to be more 2D than 1D,
and the measurement of a $g$-factor 
close to that of bulk GaAs
provides compelling evidence for the validity 
of the energy splitting technique.\cite{patel91a}
As $V_g$ is made more negative the constriction narrows,
the confinement potential becomes more rounded,
the carrier density within the channel is reduced,
and for the last few occupied subbands
the anisotropy of the electrostatic 
confinement within the channel can be described
by a saddle-point potential.\cite{lmm92} 
We, however, observe little anisotropy of the in-plane 
$g$-factor as the number of subbands is reduced,
and we believe that electron-electron interactions
are responsible for the enhancement of $g_{\parallel}$ and $g_{\perp}$.
This interpretation is in contrast to that\cite{oest95} 
in much narrower ($<100$~\AA) 1D wires,
where $\vec{k}.\vec{p}$ theory can account 
for the measured in-plane anisotropy of the $g$-factor.

In Figs.~\ref{f:2} and~\ref{f:3} we present 
transconductance ($dG/dV_g$) and conductance traces 
that show the behavior of the {\em 0.7 structure} in a 
strong in-plane magnetic field.
Figure~\ref{f:2} shows transconductance data
obtained by numerical differentiation of conductance sweeps $G(V_g)$ 
measured in fields of $B_{\parallel}=0$ to 13~T.
Due to the presence of the {\em 0.7 structure} at $B_{\parallel}=0$,
there is a satellite peak, marked with a star ($\ast$),
on the right hand side of the main transconductance peak 
that accompanies the transition from $G=0$ to $2e^2/h$.
As $B_{\parallel}$ is increased in steps of 1~T,
the satellite peak grows in intensity and the two peaks separate.
At $B_{\parallel}=13$~T the two peaks 
have roughly equal integrated areas, 
and the zero between the two transconductance peaks corresponds to
the spin-split conductance plateau at $e^2/h$. 
As a function of $B_{\parallel}$ there is a parabolic shift 
of both transconductance peaks to more positive gate voltage;
this is also observed for higher 1D subbands
and can be attributed to a diamagnetic shift of the
bottom of the 2D subband edge.\cite{stern68}
Similar parabolic variations of 
the position of spin-up and spin-down
Coulomb blockade peaks 
have been observed in a quantum dot device.\cite{weis94}

The Fig.~\ref{f:2} inset shows the gate voltage separation $\delta
V_g$ of these two transconductance peaks in positive and negative
$B_{\parallel}$.  The splitting $\delta V_g$ is linear in
$B_{\parallel}$, and the value $\delta V_g=0.035~V$ at
$B_{\parallel}=0$ which can be interpreted as a zero-field
spin-splitting with an estimated energy of $\Delta E \approx 1$~meV.
We measure both $\Delta E$ and the $g$-factors of the $n=1$ subband
from the relation $eV_{sd} = 2 g_{\parallel} \mu_B B S + \Delta E$,
rather than Eq.~\ref{energy}.  We have also measured linear Zeeman
splittings of the higher 1D subbands.

Figure~\ref{f:3} shows more clearly, in conductance,
the evolution of the {\em 0.7 structure} as 
$B_{\parallel}$ is increased in steps of 1~T.
The left hand trace at $B_{\parallel}=0$~T, 
shows a clear structure close to $0.7(2e^2/h)$,
which by 9~T has shifted down to $e^2/h$.
Figure~\ref{f:3} also shows that the structure at $0.7(2e^2/h)$ 
is not replicated at $0.7(e^2/h)$ 
when the spin degeneracy is removed at high $B_{\parallel}$,
evidence that the {\em 0.7 structure} is not a transmission effect.

Figure~\ref{f:4} shows the temperature 
dependence of the last conductance step.
As the temperature is increased from 0.07~K to 1.5~K the
definition of the plateau at $2e^2/h$ becomes weaker,
whereas the {\em 0.7 structure} strengthens 
and becomes flatter. 
The {\em 0.7 structure} is observable even at 4.2~K (not shown)
when all the quantized plateaux have disappeared.
At present we are unable to 
explain this unusual temperature behavior,
though it might imply that the {\em 0.7 structure} 
is more sensitive to 
localization than the 1D plateaux.
The behavior also provides evidence that 
the {\em 0.7 structure} is not due to an impurity,
for example, Coulomb charging would be important 
if there was ``puddling'' of the electrons close to pinch-off,
but such effects would show a weakening with increasing temperature.
Further evidence for the absence of impurity effects comes from
measurements where the channel is laterally shifted by $\pm0.04~\mu$m 
(using the technique described in Ref.~\onlinecite{williamson90});
there is little movement of the {\em 0.7 structure}
and no degradation of the higher index quantized conductance plateaux.

Due to the lack of inversion symmetry and 
the presence of interface electric fields,
zero-field spin-splitting can be present in GaAs/AlGaAs heterostructures.
Such mechanisms will simply lift the spin degeneracy of the subbands, 
and for the last subband this will give 
rise to a spin-split plateau at $0.5(2e^2/h)$ 
rather than at $0.7(2e^2/h)$.
However, it is expected\cite{glazman89} that the
energy splitting will be too small ($\sim10^{-2}$~K)
to be important in our devices.

In conclusion, we have measured the in-plane $g$-factors 
for the twenty-six subbands in a very clean 1D constriction.
In the wide channel limit we have 
demonstrated a transport technique to measure 
the $g$-factor of bulk GaAs.
There is little anisotropy of the $g$-factors in the constriction,
and the enhancement of $g_{\parallel}$ and $g_{\perp}$
as the subband index decreases is indicative that many-body
effects are important as the constriction becomes narrower.
We have presented evidence that the {\em 0.7 structure} is seen
in high mobility 1D electron gases, and is not related to impurities. 
We cannot explain the origin of this new structure,
though in a strong in-plane magnetic field we have shown that 
the {\em 0.7 structure} evolves
into the spin-split plateau at $e^2/h$;
therefore, we speculate that there is a possible spin polarization
of the 1D electron gas in zero magnetic field. 

We acknowledge useful discussions with K.-F. Berggren and A. V. Khaetskii.
We thank the Engineering and Physical Sciences 
Research Council (UK) for supporting this work. 
KJT acknowledges support from the 
Association of Commonwealth Universities,
JTN acknowledges support from the Isaac Newton Trust,
and DAR acknowledges support 
from Toshiba Cambridge Research Centre.

\begin{figure}
\caption{I. Gate voltage $G(V_g)$ characteristics showing 20 
conductance plateaux quantized in units of $2e^2/h$.
II. The gate characteristics (offset by 0.3~V for clarity)
in a magnetic field of 11~T.
{\it Insets}: (a) Detail of the structure at $0.7(2e^2/h)$.
(b) The $g$-factors as a function of subband index, 
as obtained from the Zeeman splitting at 8.2~T.}
\label{f:1}
\end{figure}

\begin{figure}
\caption{Transconductance traces $dG/dV_g$ of the 
transition between $G=0$ and $2e^2/h$
as a function of $B_{\parallel}$.
The traces have been vertically offset for clarity.
The inset shows the gate voltage splitting
$\delta V_g$  of the transconductance peak positions
as a function of $B_{\parallel}$.}
\label{f:2}
\end{figure}

\begin{figure}
\caption{The evolution of the structure at $0.7(2e^2/h)$ 
into a step at $e^2/h$
in a parallel magnetic field $B_{\parallel}=0-13$~T,
in steps of 1~T.
For clarity successive traces have been
horizontally offset by 0.015~V.}
\label{f:3}
\end{figure}

\end{document}